# Handover analysis of the Improved Phantom Cells

Pouya Sharifi [1],Hamid Shahrokh Shahraki[2]

[1,2]*Department of Electrical Engineering, University of Kashan, Kashan, Iran.*

[1] sharifi.pouya70@yahoo.com

[2] shahrokh@kashanu.ac.ir

*Abstract*— **Improved Phantom cell is a new scenario which has been introduced recently to enhance the capacity of Heterogeneous Networks (HetNets). The main trait of this scenario is that, besides maximizing the total network capacity in both indoor and outdoor environments, it claims to reduce the handover number compared to the conventional scenarios. In this paper, by a comprehensive review of the Improved Phantom cells' structure, an appropriate algorithm will be introduced for the handover procedure of this scenario. To reduce the number of handover in the proposed algorithm, various parameters such as the received Signal to Interference plus Noise Ratio (SINR) at the user equipment (UE), user's access conditions to the phantom cells, and user's staying time in the target cell based on its velocity, has been considered. Theoretical analyses and simulation results show that applying the suggested algorithm the improved phantom cell structure has a much better performance than conventional HetNets in terms of the number of handover.**

## 1. INTRODUCTION

Due to the rapid growth of wireless communication devices, construction of a new cellular structure that can meet the future high traffic demands seems necessary. Heterogeneous Networks are one the best structures proposed for this goal. Heterogeneous cellular networks are networks where a large number of small cells with low transmitter power are distributed in previous conventional Macrocells[1-2].

The mentioned small cells are known with different expression in the literatures, according to their applications. They are called Femtocells in indoor applications and known as Macro assisted small cells in outdoor ones [3-4]. Using small cells will bring about benefits such as increasing network capacity and coverage area as well as their quick and easy installation.

Nevertheless, one of the most important issues in applying HetNets is providing required quality of service (QOS) for moving users with an appropriate connection management scheme. Because of the small radius of the assisted small cells, a given mobile user may change several cells during one connection which leads to numerous number of handover [5].

In this regard, a lot of algorithms have been proposed to reduce the number of handover in HetNets which used various criteria. For instance, [6] has offered an algorithm to reduce the number of handover by taking into account the Received Signal Strength(RSS) from Macrocell and target small cell base stations(BTSs). [7] has completed this algorithm by combining RSS and measured path loss and shown that, the combined criteria offers better performance.

In another category of references in order to reduce the handover probability, handover process is done taking into account the user's velocity and its estimated staying time (dwell time) in a given cell. For example, the algorithm proposed in [8], has reduced the number of handover for users with high speed.

[9] and [10] combined different criteria and by defining a coast function suggested suitable algorithms for reducing handover rate. Among the parameters considered in the cost function, received SINR, users' velocity, available bandwidth, and small cell access mode can be noted. Reference [11] has provided a comprehensive review of the literatures related to the Femtocells handover algorithms.

On the other hand, [12] has proposed a new cellular structure called phantom cells to reduce handover in outdoor applications. In this structure, two different frequency bands are used where Macrocells use lower frequency band and phantom cells use higher frequency bands for data transmission. Moreover, in this structure, control channels' management of phantom cell is done by the covering Macrocell to reduce the number of handover and signaling overhead.

Recently, a completed model for phantom cells has introduced to maximize the throughput of these networks using carrier aggregation capability [13-14]. Moreover, the introduced structure has the advantage of being used both indoor and outdoor environments.

In this paper, by a comprehensive review of handover procedure in the improved phantom cell network, it will be shown that this structure has priority over phantom cells given the removal of handover from phantom cell to Macrocell. Besides, a proper algorithm will be suggested for cell association and handover of this network. In the proposed algorithm, in addition to received SINR form BTSs of phantom cells and Macrocells, access mode of phantom cells, and users' staying time in a phantom cell have also been taken into account. Simulation results based on real parameters confirm the suitability of the proposed handover algorithms for applying in both indoor and outdoor environments.

This paper is organized as follows. Section 2, describes the improved phantom cell structure and explains handover procedure in this network. In section 3, a suitable algorithm is presented for cell association and initial user assignment. In section 4, by exploring different types of handover in the related scenario, a proper handover algorithm is suggested. Probability analysis of the proposed algorithm is explained analytically in section 5. Performance evaluation of the proposed algorithms is verified by simulation results in Section 6 and finally, Section 7 concludes the paper.

## 2. IMPROVED PHANTOM CELL ARCHITECTURE

The overall structure of the improved phantom cell network is shown in Fig.1. As can be seen, each Macrocell contains several phantom cells. Macrocells are conventional telecommunication cells with relatively wide coverage area that, use frequency band F1 to serve their users. Phantom cells have individual base stations capable of serving their users through two frequency bands F1 and F2. In fact, phantom cell users are allowed to reuse frequency band F1 with Macrocell users considering interference constraint. In order to data synchronization and handover management, all phantom cell base stations are connected to the covering Macrocell's base station via backhaul link.
Note that, in this research we discuss about data signaling handover procedure of the network since, in the proposed structure all control signaling are managed by Macrocell base station.

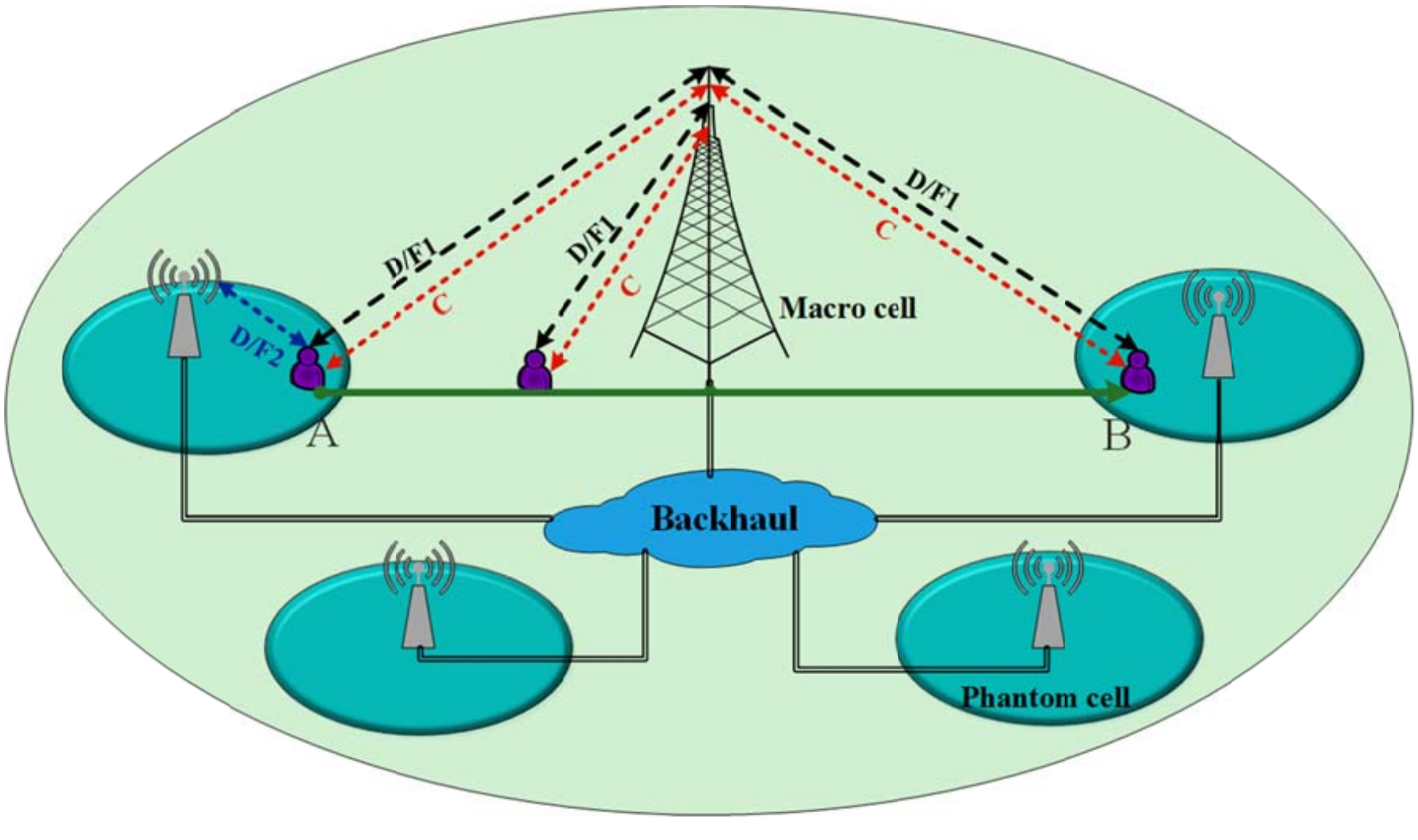

*Fig. 1-Improved Phantom cell based HetNet, where each user may receive data from frequency band F1 or both F1 and F2 depending on its position.*

It assumed that total number of Macrocells equals to $M$ and each Macrocell contains $N$ phantom cell. If the whole base stations located in the m-th Macrocell are indexed as $A = \{0,1, 2, \dots, P\}$ where $0$ corresponds to the Macrocell and the others are phantom cells' index, then the received SINR related to i-th user located in n-th cell of this Macrocell is as follows:

$$SINR_{m,n}^{i} = \frac{h_{m,n}^{i} p_{m,n}^{i}}{I_{m,n}^{i} + \sigma^{2}} \quad (1)$$

Where $p_{m,n}^{i}$ is the transmit power of n-th cell's BTS, $h_{m,n}^{i}$ denotes channel gain between n-th transmitter and i-th user's receiver, $\sigma^2$ is the additive white Gaussian noise variance, and $I_{m,n}^{i}$ represents the interference from signals of other BTSs in the given Macrocell which calculated as:

$$I_{m,n}^{i} = \sum_{j \in A \setminus \{n\}} h_{m,j}^{i} p_{m,j}^{i} \quad (2)$$

Let us assume that Macrocell and phantom cells' coverage area are circular with $R$ and $r$ radius, respectively. It is assumed that there are $U$ users uniformly distributed in the considered zone and are moving at the constant speed $V$. Two modes of access are defined for each phantom cell: open access mode, closed access mode. In open access mode, every user is allowed to access and use the phantom cell service, whereas in closed access mode only specific users are allowed to connect to the phantom cell's base station.

Accordingly, if a given user moving from point A to point B during to its transmission (see Fig.1), different statuses may be happened. It is assumed that the user's connection with Macrocells is not drop and it receives data through frequency band F1, continually. Therefore, if the given user goes through a phantom cell throughout its movement, it may connect to the phantom cell's BTS or ignore connection. It depends on the phantom cell's access mode's type, free channels situation, received SINR level and also user's velocity.

## 3. CELL ASSOCIATION

In order to optimize handover process in wireless networks, usually this process is divided into two stages. At the first stage, it is assumed that all users have been distributed as fixed in the considered geographical area. In other words, mobility issue is not considered at this stage. This stage is known as cell association.

In the cell association process, each user is assigned to the best BTS according to the considered criteria. As regards that the assumed scenario consists of two cell types (Macro and phantom) and each user can receive service from the BTSs of both cells, the cell association process should be investigated separately for both types.

At first, each user is assigned to be the best corresponding Macrocell according to the received SINR criteria. Thus, each user finds the Macrocell with the highest SINR and if it has a free channel, connects to it. If the selected

Macrocell's BTS does not have any free channel for connection, this Macrocell is removed from the list and the Macrocell with next high-priorities is checked for connection. This process continues until the intended user connects to a Macrocell.

Cell association process for phantom cells is the same with the difference that in this case, BTS's access mode should also be considered. As already mentioned each phantom cell may have open or closed access mode. If the selected phantom cell's BTS is in closed access mode for the intended user, this cell will be removed and the next high-priority ones are investigated for connection. The proposed algorithm for cell association process is summarized in Algorithm.1.

| Algorithm 1: Cell Association |
|---|
| Initialize $\eta_{m-th}$, $\eta_{f-th}$, U, M, N, $N_{m-MAX}$, $N_{ph-MAX}$ |
| $K(j) = 0 \quad \forall j = 1, \dots, M(N+1)$ |
| for i=1:U |
|   $s_i = \{j/\eta_{i,j} > \eta_{m-th} \forall j = 1, \dots, M\}$ |
|   while $s_i$ is not empty |
|     $j^* = \{j/\max \eta_{i,j} \forall j \in s_i\}$ |
|     if $K(j^*) < N_{m-MAX}$ |
|       $c_{i,j^*} = 1$ |
|       $K(j^*) = K(j^*) + 1$ |
|       empty $s_i$ |
|     else |
|       Remove BS $j^*$ from $s_i$ |
|     end if |
|   end while |
|   $v_i = \{j/\eta_{i,j} > \eta_{f-th} \forall j = M+1, \dots, M(N+1)\}$ |
|   while $v_i$ is not empty |
|     $j^* = \{j/\max \eta_{i,j} \forall j \in v_i\}$ |
|     if $a_{i,j^*} = 1 \& K(j^*) < N_{ph-MAX}$ |
|       $c_{i,j^*} = 1$ |
|       $K(j^*) = K(j^*) + 1$ |
|       empty $v_i$ |
|     else |
|       Remove BS $j^*$ from $v_i$ |
|     end if |
|   end while |
| end for |

In the above algorithm, $K(j)$ denotes the number of active users assigned to the j-th BTS and $c_{i,j}$ represents the connection status of i-th user to the j-th base station. So if $c_{i,j} = 1$, it means that i-th user is connected to j-th base station and $c_{i,j} = 0$, has reverse meaning. $\eta_{i,j}$ is the received SINR related to the connection between i-th user and j-th BTS and $a_{i,j}$ is used to show the access mode status of phantom cell base stations. $s_i$ and $v_i$ are the set of Macrocells and phantom cells whose received SINR is higher than a defined threshold. Finally, $N_{m-MAX}$ and $N_{f-MAX}$ denote the total number of serving channel in Macrocells and phantom cells, respectively.

## 4. HANDOVER MANAGEMENT

In this section, second stage of the handover procedure will described by taken into account the mobility of users. The users' mobility may cause two events. First, the user may get out of the geographical boundaries of previous cell and enter the coverage area of a new cell. Something else that may happen is that the received SINR be lower than the threshold value due to changes in the surrounding environment and fading effects.

Based on the above factors, there will be three types of handover in the proposed scenario:

- Macrocell to Macrocell handover
- Phantom cell to phantom cell handover
- Macrocell to phantom cell handover

Note that since it is assumed that the user always has its connection with Macrocell's BTS there will no phantom cell to Macrocell handover.

Generally, the conventional handover algorithms can be classified into two main categories. The first one includes algorithms where decision is made based on the received SINR. As mentioned before, the received SINR change is not implies the cell changes absolutely and it may be due to the surrounding environment changes and fading effects. Therefore, decisions based on this category may lead to additional and unnecessary handovers.

In the second type, the optimal decision is made based on estimating average staying time in a cell and average connection time considering the user's speed. In this type, parameters such as cell size and small cells' density will have an impact on the number of handovers. It should be noted that, despite the changes of user's cell in term of geographical location, the previous cell may have sufficient SINR for servicing the user. On the other hand, the user's velocity may be so high that its stay time in the next small cell be very short and connection to the new BTS may not be cost effective. Therefore, decisions based on second category may also leads to unnecessary handovers.

In this paper by combining both criteria an appropriate handover algorithm is proposed for the improved phantom cell scenario. The details of the proposed algorithm are presented in Algorithm.2.

| Algorithm 2: Handover |
|---|
| Initialize $\eta_{m-th}$, $\eta_{ph-th}$,U,M,N,K,$c$,$a$,$H_m$ , $H_{ph}$ ,$N$,$N_{m-MAX}$, $N_{ph-MAX}$<br>for i=1:U<br>  for j=1:M<br>    if $\eta_{i,j} < \eta_{m-th}$ & $c_{i,j} = 1$<br>      $s_i = \{j/\eta_{i,j} > \eta_{m-th} \ \forall j = 1, \dots, M\}$<br>      while $s_i$ is not empty<br>        j* $= \{$j/max $\eta_{i,j} \forall j \in s_i\}$<br>        if ( $\eta_{i,j^*} - \eta_{i,j} > H_m$ & $K(j^*) < N_{m-MAX}$ )<br>          $c_{i,j^*} = 1$<br>          $K(j^*) = K(j^*) + 1$<br>          $c_{i,j} = 0$<br>          $K(j) = K(j) - 1$<br>          empty $s_i$<br>        else<br>          Remove BS j* from $s_i$<br>        end if<br>      end while<br>    end if<br>  end for<br>  for j=M+1: $M(N+1)$<br>    if $\eta_{i,j} < \eta_{ph-th}$ & $c_{i,j} = 1$<br>      $v_i = \{j/\eta_{i,j} > \eta_{ph-th} \forall j = M, \dots, M(N+1)\}$<br>      while $v_i$ is not empty<br>        j* $= \{$j/max $\eta_{i,j} \forall j \in v_i\}$<br>        if ( $\eta_{i,j^*} - \eta_{i,j} > H_{ph}$ & $a_{i,j^*} = 1$ & $K(j^*) < N_{ph-MAX}$<br>            & $T_{dwell} \geq T_{expected}$<br>          $c_{i,j^*} = 1$<br>          $K(j^*) = K(j^*) + 1$<br>          $c_{i,j} = 0$<br>          $K(j) = K(j) - 1$<br>          empty $v_i$<br>        else<br>          Remove BS j* from $v_i$<br>        end if<br>      end while<br>      if $c_{i,j} = 1$<br>        $c_{i,j} = 0$<br>        $K(j) = K(j) - 1$<br>      end if<br>    end if<br>  end for |

41: if $c_{i,j} = 0 \; \forall j = M+1, \dots, M(N+1)$<br>
42: $v_i = \{ j / \eta_{i,j} > \eta_{ph-th} \forall j = M+1, \dots, M(N+1) \}$<br>
43: while $v_i$ is not empty<br>
44: $j^* = \{ j / \max \eta_{i,j} \forall j \in v_i \}$<br>
45: if ( $a_{i,j^*} = 1$ & $K(j^*) < N_{ph-MAX}$ & $T_{dwell} \geq T_{expected}$)<br>
46: $c_{i,j^*} = 1$<br>
47: $K(j^*) = K(j^*) + 1$<br>
48: empty $v_i$<br>
49: else<br>
50: Remove BS $j^*$ from $v_i$<br>
51: end if<br>
52: end while<br>
53: end if<br>
54: end for

As already explained, three types of handover may occur in the defined scenario, each of which will describe in detail in the following.

- **Macrocell to Macrocell handover**

This type of handover happens between Macrocell BTSs which use the main frequency band, i.e. F1. If the server's SINR becomes less than the predetermined value, Macrocell to Macrocell handover process begins. Thus, the user considers a list of macro cells whose SINR value is more than the threshold. BTS that has the maximum SINR value is selected as the target candidate. If the target Macrocell SINR value is greater than handover hysteresis value, and the target cell has free channel, handover is done. Otherwise, this Macrocell is removed from the list, and the process is repeated for other Macrocell until the user is connected to a new Macrocell's BTS.

- **Phantom cell to phantom cell handover**

This type of handover, which is related to the phantom cell users on frequency band F2, is addressed in the second part of the algorithm (Line 20). General principles of management of this type of handover are the same as before with the differences listed below.

After determining the list of target phantom cells based on SINR and selecting the phantom cell with highest priority, the following conditions must be met for handover to take place:

- SINR of the target cell must be larger than the SINR of serving cell, as large as hysteresis value.
- Target cell must have free channel for serving.
- The user must have access to the target cell.(open access mode)

- The user's estimated stay time in the coverage area of the target phantom cell (dwell time) must be greater than specific value (expected connection time). In the other word, if the dwell time is not so enough the new connection is not cost effective and the user ignores new connection.

If the intended phantom cell does not have the above conditions, it is removed from the target phantom cell's set and another phantom cell that has the highest priority is selected. This process continues until the user connects to a phantom cell or phantom cell's set be empty. Obviously, if the user fails to connect any phantom cell, its connection through frequency band F2 will be cut off.

- **Macrocell to phantom cell handover**

As mentioned above, the user's connection to the previous phantom cell may break due to moving and it may be unable to connect another phantom cell's BTS. In this case, until the reconnection to a new phantom cell's BTS, the user receives data only from Macrocell. With reconnection of the user to a new phantom cell, frequency band F2 is reactivated, which is considered as macro cell to phantom cell handover. The details of managing this handover are addressed in the third part of the algorithm (line 41).

## 5. PROBABILTY ANALYSIS

In this section, we will explore handover problem analytically by modeling different states that may accrue to a user. To do so, a given user is considered in a Macrocell region and different states are shown based on Markov chain approach. As shown in Fig.2, in the assumed Markov model each user may be in one of the following states.

- The user is in the coverage area of Macrocell and receives data only from Macrocell's BTS.
- The user is in the coverage area of a phantom cell and receives data from Macrocell's BTS.
- The user is in the coverage area of a phantom cell and receives data from both phantom cell and Macrocell BTSs simultaneously.

The probability of occurrence of any of the above events is shown with $P_{S1}$, $P_{S2}$ and $P_{S3}$. To examine transition states based on the proposed handover algorithm, we model the user status in discrete time intervals. Thus, $P_{S1}[k]$ represents the probability of being in the state $S1$ at the k-th moment. State transition matrix of Fig.2, is as follows.

$$\begin{bmatrix} P_{S1}[k] \\ P_{S2}[k] \\ P_{S3}[k] \end{bmatrix} = \begin{pmatrix} P_{11}[k] & P_{12}[k] & P_{13}[k] \\ P_{21}[k] & P_{22}[k] & P_{23}[k] \\ P_{31}[k] & P_{32}[k] & P_{33}[k] \end{pmatrix} \begin{bmatrix} P_{S1}[k-1] \\ P_{S2}[k-1] \\ P_{S3}[k-1] \end{bmatrix} \quad (3)$$

Where, $P_{ij}$ is the probability of transition from state $j$ to state $i$.

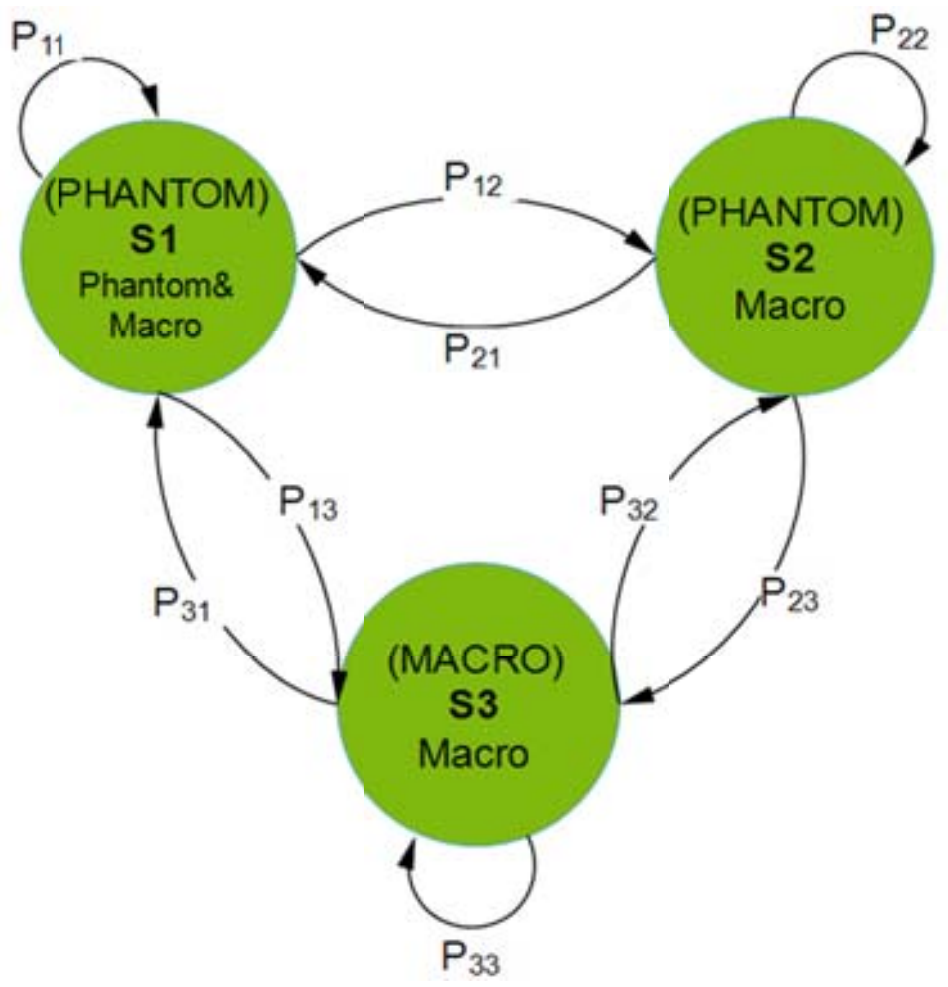

*Fig.2-Markov chain state transition*

In order to explore the probability of handover in the intended network, we examine the probability of transition for one state. The probable analysis related to the other states will be similar.

Let us assume that the target user's state is $S2$ at the time interval $[k-1]$ and switched to state $S1$ at the k-th moment as a result of moving. If we show the probability of this event with $P_{ho-12}$, then it can be computed as(note that this probability is equal to $P_{12}[k]$ in Eq.(3)):

$$P_{ho-12} = P(S_1[k]/S_2[k-1]) =$$

$$P\{\eta_{ph}[k] > \eta_{ph-th}, access\ mode = 1, N_{ph-MAX}, T_{dwell} \geq T_{expected}/S_2[k-1]\} =$$

$$P(\eta_{ph}[k] > \eta_{ph-th}/S_2[k-1]).P(access\ mode = 1/S_2[k-1]).$$

$$P(N < N_{ph-MAX}/S_2[k-1]).P(T_{dwell} \geq T_{expected}/S_2[k-1]) =$$

$$P(\eta_{ph}[k] > \eta_{ph-th}/S_2[k-1]).P(access\ mode = 1).P(N < N_{ph-MAX}).P(T_{dwell} \geq T_{expected}) \quad (4)$$

In order to simplify the calculations, we describe each part of the above probability, separately.

- $\boldsymbol{P(access\ mode = 1)}$

The above probability depends on close or open access class of operation of the desired phantom cells. In our analysis and simulations, this probability has been considered equal to $\frac{1}{2}$.

- $\boldsymbol{P(T_{dwell} \geq T_{expected})}$

The above probability can be rewritten as below:

$$P(T_{dwell} \geq T_{expected}) = 1 - P(T_{dwell} < T_{expected}) \quad (5)$$

Where, $P\{T_{dwell} < T_{expected}\}$ is the probability of the user's staying time in a phantom cell being smaller than its expected time of transmission. To calculate this probability, we generalize the results obtained in [15], so:

$$P(T_{dwell} < T_{expected}) = \frac{T_{expected}}{T_{expected}+T_{dwell}} \quad (6)$$

And as a result:

$$P(T_{dwell} \geq T_{expected}) = 1 - \frac{T_{expected}}{T_{expected}+T_{dwell}} = \frac{T_{dwell}}{T_{expected}+T_{dwell}} \quad (7)$$

- $\boldsymbol{P(N < N_{ph-MAX})}$

By rewriting the above equation in a similar way, we have:

$$P(N < N_{ph-MAX}) = 1 - P(N \geq N_{ph-MAX}) \quad (8)$$

In the above equation, $P\{N \geq Nmax\}$ indicates the probability that the target phantom cell's BTS has no free channel. If we assume that the total number of channels assigned to a phantom cell is $T$ and from these channels, g channels are exclusively reserved for handover process, then:

$$P(N \geq N_{ph-MAX}) = \rho^{T-g} \, \frac{\rho_h^g}{T!} \, P_0 \quad (9)$$

Where $\rho_h$ and $\rho$ define as:

$$\rho = \frac{(\lambda_n+\lambda_h)}{\mu_c} \quad (10)$$

$$\rho_h = \frac{\lambda_h}{\mu_c} \quad (11)$$

In the above equations, $\lambda_n$, $\lambda_h$ and $\frac{1}{\mu_c}$ represent the average new traffic intensity, new call arrival rate and average channel occupancy time, respectively [15]. Moreover, $P_0$ is calculated as follows:

$$P_0 = \left[\sum_{j=0}^{T-g} \frac{\rho^j}{j!} + \rho^{T-g} \sum_{j=T-g+1}^{T} \frac{\rho_h^{j-(T-g)}}{j!}\right]^{-1} \quad (12)$$

Therefore, by calculating the expression (9) based on the network traffic parameters and replace it in Eq. (8), the intended probability can be calculated.

- $\boldsymbol{P(\eta_{ph}[k] > \eta_{ph-th}/S_2[k-1])}$

Using the conditional probability rules the above probability can be calculate as:

$$P(\eta_{ph}[k] > \eta_{ph-th}/S_2[k-1]) = \frac{P(\eta_{ph}[k]>\eta_{ph-th}, S_2[k-1])}{P(S_2[k-1])} \tag{13}$$

To derive the above probability it is necessary to calculate the probability of being the user in the state $S_2$ at time interval $[k-1]$. To do so, we can write:

$$\{S_2[k-1]\} =$$

$$\{\eta_M[k-1] > \eta_{m-th}, access\ femto = 0\} \cup \{\eta_M[k-1] > \eta_{m-th}, N > N_{ph-MAX}\}$$

$$\cup \{\eta_M[k-1] > \eta_{m-th}, T_{dwell} < T_{expected}\} \cup \{\eta_M[k-1] > \eta_{m-th}, \eta_{ph}[k-1] < \eta_{ph-th}\} \tag{14}$$

Ignoring the joint probability of the events in Eq. (14) and also by taking to account the independence of the received SINR from BTSs of phantom cells and Macrocell, it can be written:

$$P(S_2[k-1]) = P(\eta_M[k-1] > \eta_{m-th}).$$

$$\{P(access\ femto = 0) + P(N > N_{ph-MAX}) + P(T_{dwell} < T_{expected}) + P(\eta_{ph}[k-1] < \eta_{ph-th})\} \tag{15}$$

The numerator of Eq. (13) can also be written as:

$$P(\eta_{ph}[k] > \eta_{ph-th}, S_2[k-1]) =$$

$$P(\eta_M[k-1] > \eta_{m-th}). \left\{ \begin{array}{l} [P(\eta_{ph}[k] > \eta_{ph-th}). P(access\ mode = 0)] \\ +[P(\eta_{ph}[k] > \eta_{ph-th}). P(N > N_{ph-MAX})] \\ +[P(\eta_{ph}[k] > \eta_{ph-th}). P(T_{dwell} < T_{expected})] \\ +P[\eta_{ph}[k] > \eta_{ph-th}, \eta_{ph}[k-1] < \eta_{ph-th}] \end{array} \right\} \tag{16}$$

Therefore, considering Equations (7) and (8), the remaining task for calculating Eq. (16) is to calculate probabilities $P\{\eta_M[k-1] > \eta_1\}$ and $P\{\eta_f[k] > \eta_{th}, \eta_{PH}[k-1] < \eta_2\}$. Given the Gaussian nature of probability density function of received SINR, these probabilities can be as follows [16].

$$P(\eta_{ph}[k-1] < \eta_{ph-th}) = Q\left(\frac{\eta_{ph-th} - \mu_{\eta(k-1)}}{\sigma_{\eta(k-1)}}\right) \tag{17}$$

$$P(\eta_{ph}[k] > \eta_{ph-th}, \eta_{ph}[k-1] < \eta_{ph-th}) = \int_{-\infty}^{\eta_{ph-th}} f_{\eta(K/K-1)}(\eta)\, Q\left(\frac{\eta_{ph-th} - \mu_{\eta(K/K-1)}}{\sigma_{\eta(K/K-1)}}\right) d\eta \tag{18}$$

Where:

$$f_{\eta(K/K-1)}(\eta) = \frac{1}{\sqrt{2\pi\sigma^2_{\eta(K/K-1)}}} . e^{-\left(\frac{1}{2\sigma^2_{\eta(k-1)}}\right)\left[\eta_f[k]-\mu_{\eta(K/K-1)}\right]^2} \quad (19)$$

$$\mu_{\eta(K/K-1)} = \mu_{\eta(K)} + \rho_{\eta\eta} \frac{\sigma_{\eta(K)}}{\sigma_{\eta(K-1)}} \left(\eta_{ph}[k-1] - \mu_{\eta(K-1)}\right) \quad (20)$$

$$\sigma^2_{\eta(K/K-1)} = \sigma^2_{\eta(K)} \left(1 - {\rho_{\eta\eta}}^2\right) \quad (21)$$

$$E\{\eta[K-1]\} = \mu_{\eta(k-1)} \qquad E\{\eta[K]\} = \mu_{\eta(k)} \quad (22)$$

$$Var\{\eta[k-1]\} = \sigma_{\eta(k-1)} \qquad Var\{\eta[k]\} = \sigma_{\eta(k)} \quad (23)$$

Finally, $\rho_{\eta\eta}$ represents the correlation coefficient between samples $\eta_{ph}[k]$ and $\eta_{ph}[k-1]$.

## 6. SIMULATION RESULTS

In this section, by simulating the proposed algorithm applied in the improved phantom cell scenario, its performance is evaluated. Both indoor and outdoor scenarios are considered in the performed simulations.

Since the considered scenario in this paper is a new scenario, it is not possible to compare its results with the results of other algorithms. However, for a relative comparison, the obtained results have been compared with the results of algorithm presented in [17].Note that in the scenario of [17], the user is just connected to one of BTSs related to Macrocell or small cell. In other words, the user is only able to receive data from one frequency band. For a fair comparison the proposed algorithm in [17], has been modified and simulated based on the current paper's criteria.

The simulations carried out in accordance with Fig.3, where in the indoor scenario, 2 macro cells are intended, each of which includes 12 phantom cells, and in the outdoor scenario, 2 macro cells are considered, each of which includes 8 phantom cells.

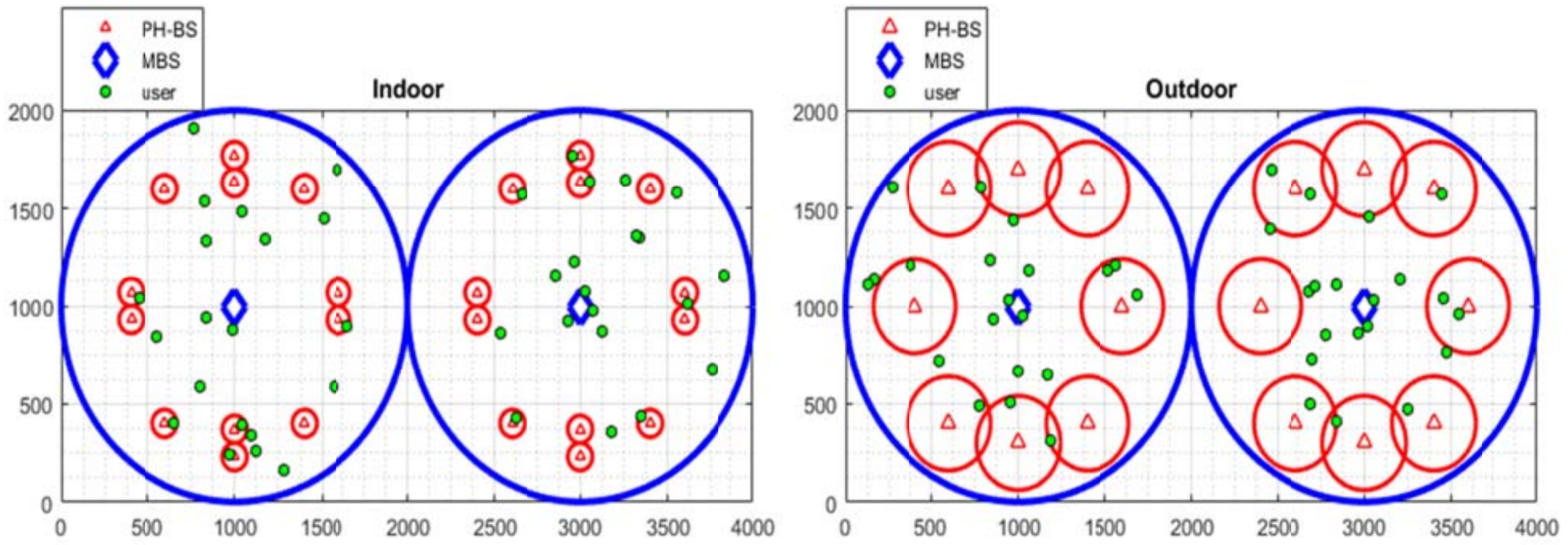

*Fig.3- the simulated network's geometry in including both outdoor and indoor cases*

Users are uniformly distributed on the coverage area of cells and each user is directly moving in a random direction. The users' speed are considered uniformly in the interval [0-4.1 m/s] and [0-8.3 m/s] for the indoor and outdoor cases, respectively. The received SINR at the user's front-end is calculated based on equations (1) and (2) where the related channel coefficients are considered based on the formulas offered in Table (1).

In this table, $d$ represents the distance of the user with its serving BTS and δ denotes the shadowing effect which is considered as a random variable with zero mean, normal distribution and variance equal to 6 $dB$. $L_{ow}$ is the indicator of free space path loss which is considered equal to 10 $dB$ in the simulations. Moreover, in the indoor case, the walls' loss also has been added as $q.w$ where $q$ represents the numbers of walls and $w$ is the loss coefficient equal to 5 $dB$.

Other parameters used in the simulations are summarized in Table (1).

*Table 1- Simulation parameters*

| | Parameter | Definition |
|---|---|---|
| Outdoor | $PL_0 = PL_1 = 15.3 + 37.6\log_{10}(d) + u$ | Path loss model for Macrocell |
| | $PL_j = \max\big(15.3 + 37.6\log_{10}(d)\,, 3 + 20\log_{10}(d)\big) + L_{ow}$ | Path loss model for phantom cell |
| | N=8 | The number of phantom cells per Macrocell |
| | r=250 m | phantom cells radius |
| | $Pt_j = 31.5\ dBm$ | Transmit power of the j-th phantom cell |
| Indoor | $PL_0 = PL_1 = 15.3 + 37.6\log_{10}(d) + qw + L_{ow}$ | Path loss model for Macrocell |
| | $PL_j = 37 + 20\log_{10}(d) + qw$ | Path loss model for phantom cell |
| | N=12 | The number of phantom cells per Macrocell |
| | r=50 m | phantom cells radius |
| | $Pt_j = 23\ dBm$ | Transmit power of the j-th phantom cell |
| | M=2 | The number of Macrocells |
| | R=1000 m | Macrocells radius |
| | $Pt_0 = Pt_1 = 43\ dBm$ | Transmit power of the Macrocell |
| | $\delta = 6\ dB$ | Shadowing effects variance |
| | $\sigma^2 = -170\ dBm$ | Noise power |
| | $N_{ph-MAX} = 10$ | Maximum number of users per phantom cell |
| | $H_m = H_m$ =H=0.1 | Hysteresis Margin |
| | $T_{th} = 5\ s$ | dwell time threshold |
| | $\eta_{ph-th}$=0.45 | SINR threshold for phantom cell |
| | $\eta_{m-th}$=0.40 | SINR threshold for Macrocell |
| | $L_{ow} = 10\ dB$ | Outdoor penetration loss |
| | $w = 5dB$ | Wall partition loss |

In Fig.4, the average number of handover is shown based on the number of users. In the first part of all graphs, with increase in the number of users the number of handover rises as expected. However, as the graphs show, increasing the number of users greater than a specified value leads to reduction of average number of handover. This is due to the fact that with occupation of all phantom cells channels there will be no any free channel for handover operation and so the number of handover decreases.

By comparing the results of indoor and outdoor cases, it is observed that the number of handover in indoor environment is much more than outdoor ones. One reason is that the received SINR variations in indoor environments are more which leads to the more phantom cells link connection and disconnection.

Moreover, the results of the proposed algorithm in the current paper can be compared with the results of the algorithm introduced in [17]. As is seen, the number of handover in our algorithm is outperforming than the algorithm of [17], about 40% in indoor and 80% in outdoor applications. As mentioned through the paper, the main reason of this significant improvement is that there is no phantom cell to Macrocell handover in our algorithm.

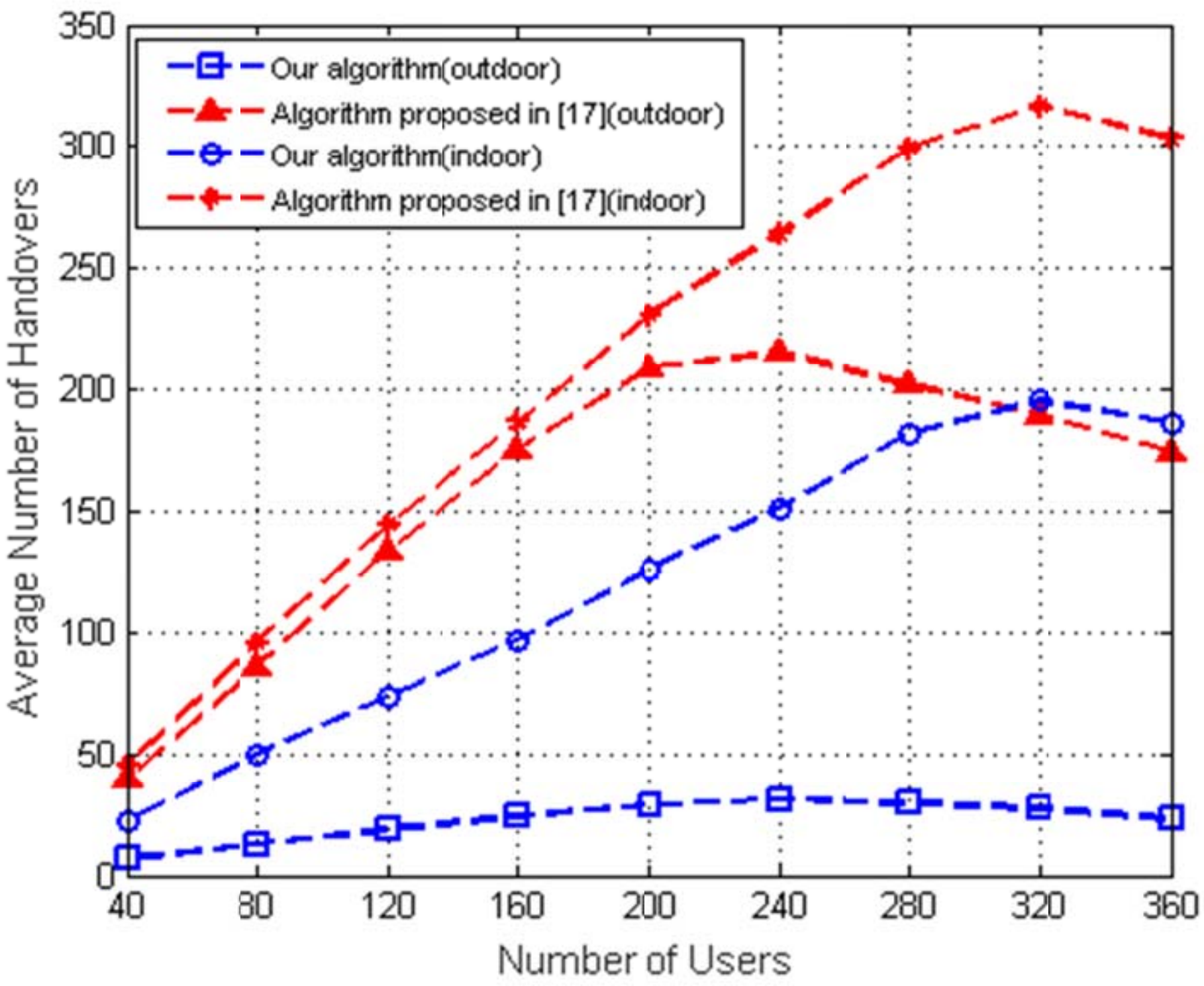

*Fig.4-Copmarision of the average number of handover based on the number of users*

In Fig.5, we have dealt with examining the effect of considering user's staying time in the proposed algorithm. As can be seen, adding this parameter in the proposed algorithm has reduced the number of handovers to about half. Furthermore, by exploring the obtained graphs it can be seen that the effect of this parameter in the outdoor case is much lower. This is because in outdoor mode the coverage area of a phantom cell is wider and therefore usually the user's staying time in the cell will be longer than the expected time of transmission. This brings about fewer phantom cells ignoring because of users staying time.

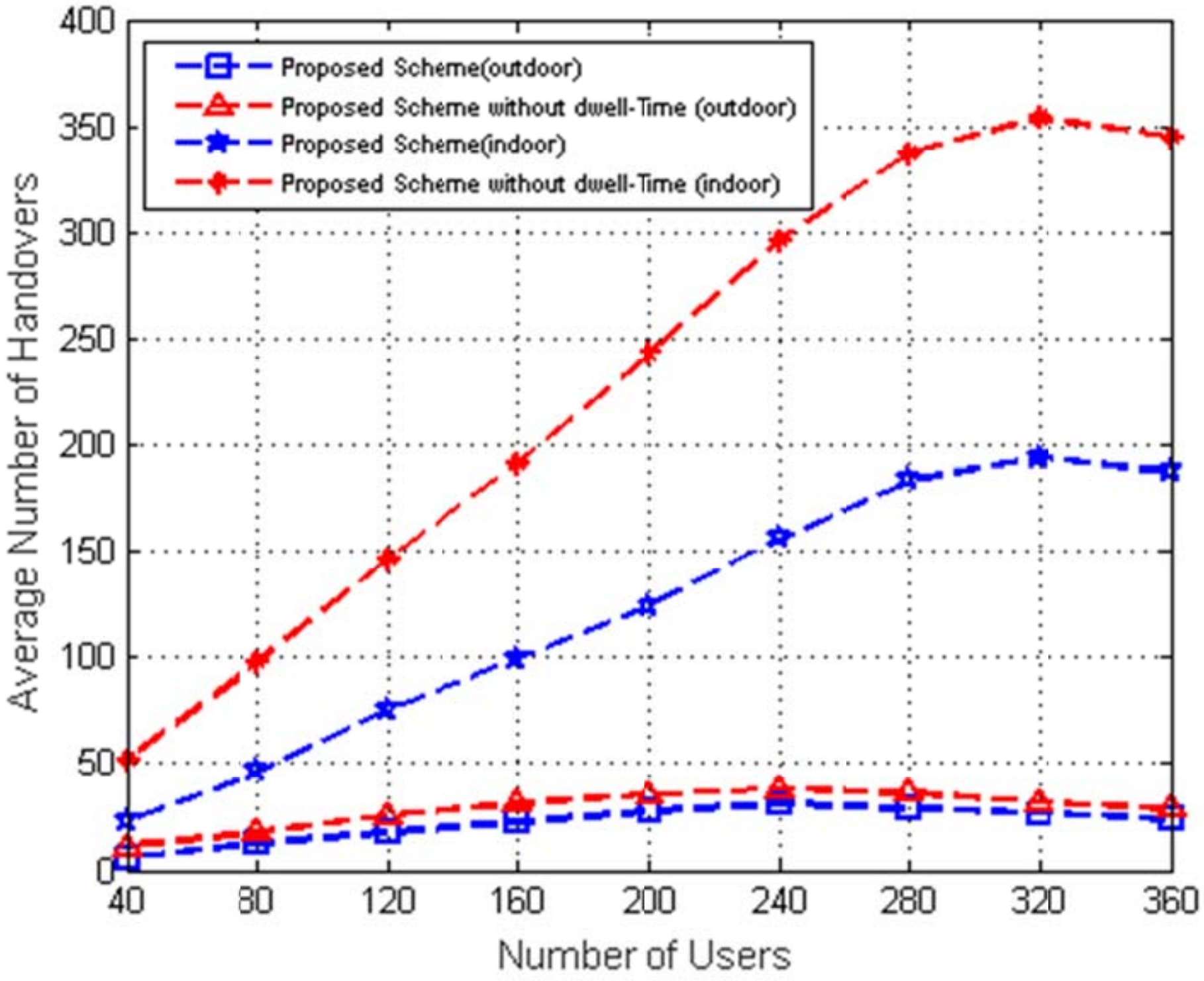

*Fig.5- Evaluating the effect of considering dwell time in the proposed algorithm*

One of the most important parameters of the proposed algorithm is the handover hysteresis value. The effect of this value on the performance of the algorithm is shown in Fig.6. As the figure shows, by increasing the value of the hysteresis, the average number of handover will be decreased. However, it should be noted that selecting the large value for the hysteresis may increase the probability of dropped connections.

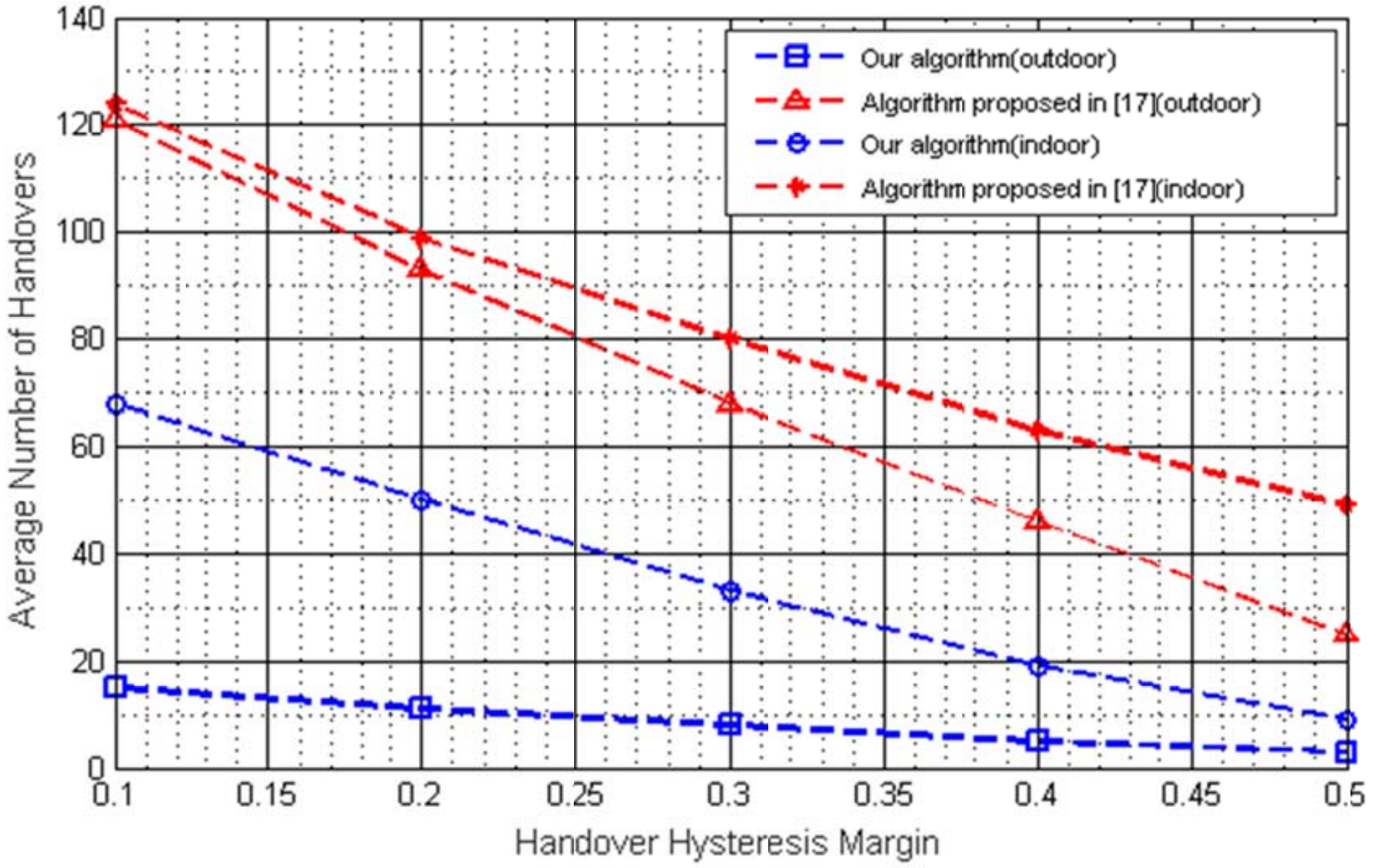

*Fig.6- Evaluating the effect of choosing Hysteresis value in the proposed algorithm*

## 7. CONCLUSIONS

In this paper the handover problem in the improved phantom cell based HetNets is investigated. In this context, the improved phantom cell scenario has been explored and the main criteria in handover decisions are introduced. In order to improve the performance of handover procedure this process is classified into two stages, known as, cell association and handover management. By investigating different aspects of each stage, a proper algorithm has been proposed to apply in each stage. Analytical description of the proposed algorithms is derived using discrete time Markov chain model. Finally, with examining the effectiveness of the proposed algorithm and comparing with other ones by numerical experiment, its dramatic improvement is proved. As a result, we can conclude that the improved phantom cell structure is outperforming to the other HetNet structures not only from throughput point of view but also from handover feature.